\documentclass[12pt]{article}
\usepackage{epsf}
\begin{document}

\begin{center}

RING DIAGRAMS AND ELECTROWEAK PHASE TRANSITION IN A MAGNETIC FIELD

\vskip 0.5cm

Vladimir Skalozub$^a$  and  Michael Bordag$^b$

\vskip 0.5cm
$^a$ Dniepropetrovsk University, Dniepropetrovsk 320625, Ukraine\\
e-mail: skalozub@ff.dsu.dp.ua 

$^b$ Institut f{\"u}r Theoretische Physik, Leipzig University,
     04109, Leipzig, Germany, e-mail: Michael.Bordag@itp.uni-leipzig.de

\end{center}

\vskip 0.5cm


\vskip 0.5cm

\begin{abstract}
Electroweak phase transition in a magnetic field is investigated within
the one-loop and ring diagram contributions to the effective potential in the minimal Standard Model.
All fundamental fermions and bosons are included with their actual values of masses and the Higgs boson mass is considered in the range $75 GeV \leq m_H \leq 115  GeV$. The effective potential is real at sufficiently high temperature. The important role of fermions and $W$-bosons in symmetry behaviour is observed. It is found that the phase transition for the field strengths $10^{23} - 10^{24}$G is of first order but the baryogenesis condition is not satisfied. The comparison with the hypermagnetic field case is done. 
\end{abstract}

1. The concept of symmetry restoration at high temperature has been intensively used in studying the evolution of the universe at its early stages. Nowadays it gives a possibility to investigate various problems of cosmology and particle physics \cite{DK},\cite{ADL}. In particular, the type of the electroweak phase transition and hence the further 
evolution of the universe depend on the Higgs boson mass $m_H$.  Most 
investigations of the electroweak (EW) phase transition have  included into 
consideration  high temperature as the main environment \cite{ADL}, \cite{She}. But 
in recent years  cogent arguments following from different approaches in 
favour of the 
presence of strong  magnetic fields  at that stage have appeared 
\cite{Vac}, \cite{Bra} (for recent review see \cite{Enq}). So, the phase 
transition at high temperature and strong fields has to be of interest. Moreover, at 
present time when  masses of all fundamental particles, except $m_{H}$, are 
known it is desirable to investigate in details the phase transition as the 
function of this parameter. 

One of the possibilities to have  strong  magnetic fields in the  EW
phase transition epoch was discussed by Vachaspati \cite{Vac}.
From his analysis it follows that under very general conditions the fields 
$H \sim T^2_i$ in the patches of sub-horizon scales can be generated during a large 
class of  grand unified transitions \cite{Bra},\cite{Enq}, where $T_i$ is transition 
temperature. The second one is the formation of the Savvidy  vacuum magnetic state at 
high temperature ($H \sim g T^2, g $ is gauge coupling constant) 
\cite{Cabk},\cite{SVZ},\cite{EnO},\cite{SkT}. In latter case only the abelian field 
configurations could arise spontaneously since they are sourceless.  For many 
problems of cosmology it is important to estimate the field strengths presented, but 
it is difficult to realise that without detailed investigations within specific models. 
Usually, only one type of fields is considered. Therefore, results obtained in such a 
way give an upper estimate of the field. This remark is relevant to our  analysis. 

Various aspects of the phase transitions in magnetic fields at high temperature have 
been investigated by many authors \cite{SVZ}-\cite{Rez}. In Ref.\cite{Riot} the influence of magnetic field on the sphalerons and the possible consequences of that for the EW phase transition were considered.   These studies are 
concentrated mainly on the influence of the boson fields only. But due to a rather heavy 
$t$-quark mass, $m_t \simeq 175 GeV,$ the influence of fermion sector increases. This is the case in strong magnetic fields even at low temperature due to the presence in the one-loop effective potential (EP) of the term $\sim m_f^2 eH$, where $m_f$ is fermion mass. Moreover, at high temperature not only heavy but also light fermions are important, as it follows from the term $\sim H^2 log T/m_f$  of the EP which significantly influences the EP curve in the broken phase. Actually,  it will be shown below that the strong magnetic fields affect essentially the 
phase transition dynamics. Another aspect of the phase transition, which also was 
not investigated but plays an important role, is the influence of the so-called ring 
diagrams at high temperature and strong field.  At zero field it was considered 
in Refs.\cite{Tak},\cite{Car}  where  their importance for 
determining of the type of the phase transition has been shown. In the latter paper the t-quark mass was 
taken of order $110 $Gev. So, taking into account present day data, it should be considered as 
a qualitative estimate of the  effect of ring diagrams even for zero-field case. 

The aim of the present paper is to  investigate the  EW phase transition at high 
temperature and constant strong magnetic fields $H$.  We consider  the case when 
the magnetic field  is present in both, the restored and broken phases. This scenario may be 
realized in nature when the field is generated due to the Savvidy mechanism at a GUT 
scale. We calculate and investigate the one-loop EP  and the 
contributions of ring diagrams.  We include  all 
bosons and fermions with their actual masses. So, the mass $m_{H}$ remains  the only free parameter. Taking into account the present day experimental limit $m_H \ge 90 GeV$, we consider the range of the mass values $75 GeV \le m_H \le 115 GeV.$ The lower bound chosen corresponds to the values of the mass when perturbative methods give reliable results. For heavier masses it should be considered as an estimate. But we belive that the effects of strong fields being included exactly are not very sensitive to the change of the mass range investigated. With these approximation adopted we observed that for weak magnetic fields the phase transition is of second order or of weak first order. The increase of the field strength make it stronger first order. But even for the field strengths $H \sim 10^{23} - 10^{24} G$ the baryogenesis does not survive in the minimal Standard Model (SM).
 
It will be important for what follows to remember recent results on
 obeservation of the gluon magnetic mass in lattice simulations that
 was found to be of order $m_{mag} \sim g^2 T$ (as has been expected
 from nonperturbative calculatios in quantum field theory \cite{Kal}, \cite{Buch}). The mass 
screens  the nonabelian component of magnetic fields at distances $l > l_m \sim (g^2 
T)^{-1}$ but inside the space region $l < l_m$  it may
 exist. Since the typical order  of particle masses at high temperature is $M  \sim gT$ ,  
the spectrum of charged particles is formed at  the space range  of Larmor's radius 
$r_L \sim (gH)^{-1/2}$ and the magnetic field strength  generated at high temperature has the order $(gH)^{1/2}\sim g^2 T  $\cite{SVZ}, \cite{SkT} the field is able to affect all the processes 
at high temperatures. The latter fact was not taken into account in a number of 
investigations of the EW phase transition. In particular, in recent papers \cite{Elmp},\cite{Pers} ( as in 
Ref. \cite{EnO}) the field strength generated at finite temperature was erroneously estimated 
as coinciding with that at zero temperature. Hence,  it has been concluded   that the magnetic 
fields could not be spontaneously generated at all (because  for weak fields generated in the 
vacuum the Larmor radius is larger then the inverse magnetic mass and such fields must 
be screened).

In papers \cite{Shap1}, \cite{Elm1}, \cite{Shap2} the influence of the 
hypermagnetic field on the EW phase transition has been investigated. In  Ref. 
\cite{Shap1} the EP 
was computed in a tree approximation and the result that the presence of $H_Y$
makes the week first-order phase transition stronger has been derived.  In Ref. 
\cite{Elm1} the temperature dependent part of the EP was calculated in one-loop order  whereas the field has also been taking into account at tree level. By 
investigating the EP these authors  came to the conclusion that the hypermagnetic 
field induces strongly first order EW phase transition. Moreover, they found 
that for  $H_Y  > 0.3 - 0.5 T^2$, where $T$ is the transition temperature, the 
standard baryogenesis survives. However, we  would like to 
note that for the week first-order phase transition the fluctuations are essential,  
the one-loop approximation to the EP is not sufficient and the correlation 
corrections must be included \cite{Car}, \cite{Kal}. 
In Refs.\cite{Shap2}, \cite{Lain} this problem has been investigated by lattice 
simulations.  We  will compare that results with the ones presented here in the 
last section.

To make a link between studies of symmetry behaviour  in external hypermagnetic 
field and previous results for the case of usual magnetic field \cite{Rez}, 
\cite{SVZ} we note that  in the broken phase  $H_Y$ and $H$ are connected by the 
relation $H = H_{Y} \cos \theta_w$, where $\theta_w$ is the Weinberg angle. So, all 
investigations dealing with symmetry behaviour in a magnetic field at high temperature are 
relevant to the case of $H_Y$ in the respect of the form of the EP curve at different $T, 
H_Y$. The hypercharge field influences the scalar field condensate at tree level and acts to 
restore symmetry. That was the reason why it has been taken into account in the lowest 
order. But, as it will be shown below,  for strong fields and heavy $m_H$ the form of the 
EP curve in the broken phase is very sensitive to the change of the parameters. Moreover, 
it is strongly dependet on the correlation correction contributions of heavy particles. So, to 
have an adequate picture of the  EW phase transition the symmetry behaviour with rings 
included has to be investigated.
~\\~

 2.  The Standard Model  Lagrangian is well known (see, for example, 
Refs. \cite{ADL}, \cite{Sk2}). The one-loop contributions of bosons and 
fermions to the EP  at finite temperature and magnetic field have
been calculated and for detais we refer  readers to papers 
\cite{Rez},\cite{Sk2},\cite{SVZ},\cite{Elm}. Below, we consider only some 
necessary  information about that and concentrate our attention on calculation of   
the ring diagrams.

The external electromagnetic field is introduced by splitting the potential in 
two parts: $A_{\mu}= \bar{A_{\mu}} + A^{R}_{\mu} $, where $A^{R}$ describes a 
radiation field and $\bar{A} = (0,0,Hx^1,0)$ corresponds to the constant 
magnetic field directed along the third axis. We make use of the gauge-fixing
conditions \cite{Sk2}
\begin{equation} \label{3} \partial_{\mu}W^{\pm \mu} \pm ie\bar{A_{\mu}}
W^{\pm \mu} \mp i\frac{g\phi}{2\xi}\phi^{\pm} = C^{\pm}(x),
\end{equation}
\begin{equation} \label{4} \partial_{\mu}Z^{\mu} - \frac{i}{\xi'}
(g^2 + g'^2)^{1/2}\phi_{z} = C_z ,
\end{equation}
where $ e = g sin \theta_w, tan \theta_w = g'/g, \phi^{\pm}, \phi_{z}$ are the 
Goldstone fields, $\xi, \xi' $ are the gauge fixing parameters, $C^{\pm}, C_z$ 
are arbitrary functions and $\phi_c$ is a scalar condensate value. In what 
follows, all calculations will be done in the general relativistic 
renormalizable gauge (\ref{3}),(\ref{4}) and after that we  set $\xi,\xi' = 0$ 
choosing the unitary gauge.

To compute the EP $V^{(1)}$ in the background magnetic field let us introduce the 
proper time, s-representation for the Green functions
\begin{equation} G^{ab}= - i \int\limits_{0}^{\infty} ds \exp(-is {G^{-1}}^{ab})
\end{equation}
and apply the method of Ref.\cite{Cab}, allowing in a natural way to incorporate 
the temperature into this formalism. A basic formula of Ref.\cite{Cab} 
connecting the Matsubara-Green functions with the Green functions at zero 
temperature is needed,
\begin{equation} \label{5} G^{ab}_k(x,x';T) = \sum\limits_{-\infty}^{+\infty}
(-1)^{(n+[x])\sigma_k} G^{ab}_k(x-[x]\beta u, x'- n\beta u),
\end{equation}
where $G^{ab}_k $ is the corresponding function at $T=0, \beta =1/T, u =(0,0,0,1),
$ the symbol $[x]$ denotes the integer part of $x_{4}/\beta, \sigma_k = 1$ in the case 
of physical fermions and $\sigma_{k} =0$ for boson and ghost fields. The Green 
functions in the right-hand side of formula (\ref{5}) are the matrix elements
of the operators $G_k$ computed in the states $\mid x',a)$ at $T=0$, and in the 
left-hand side the operators are averaged over the states with $T \not= 0$. The 
corresponding functional spaces $U^{0}$ and $U^{T}$ are different but in the 
limit of $T \rightarrow 0~~  U^{T}$ is transformed into $U^{0}$.

The one-loop contribution to the EP is given by the expression \cite {Sch}, \cite 
{Sk2}
\begin{equation} \label{6} V^{(1)} = - \frac{1}{2} Tr\log G^{ab},
\end{equation}
where $G^{ab}$ stands for the propagators of all the quantum fields $W^{\pm}, 
\phi^{\pm},...$ in the background magnetic field $H$. In the s-representation 
the calculation of the trace can be done in accordance with formula \cite{Sch}
\begin{equation} Tr\log G^{ab} = - \int\limits_{0}^{\infty} \frac{ds}{s}
tr \exp(-is G^{-1}_{ab} ).
\nonumber
\end{equation}
Details of calculations based on the s-representation and the formula (\ref{5}) can 
be found, for example, in Refs.\cite{Cab},\cite{Rez}. The terms with $n=0$ in 
Eqs.(\ref{5}), (\ref{6}) give  zero temperature contributions to Green's 
functions and effective potential $V^{(1)}$, respectively. They are the only 
terms possessing divergences. To eliminate them and uniquely fix the potential 
we use the following renormalization conditions at $H,T = 0$\cite{Rez}:
\begin{equation} \label{7} \frac{\partial^2 V(\phi,H)}{\partial H^2}\mid_{H=0,
\phi=\delta(0)} = \frac{1}{2} ,
\end{equation}
\begin{equation} \label{8} \frac{\partial V(\phi,H)}{\partial \phi}\mid_{H=0,
\phi=\delta(0)} = 0,
\end{equation}
\begin{equation} \label{9} \frac{\partial^2 V(\phi,H)}{\partial \phi^2}
\mid_{H=0,\phi=\delta(0)} = \mid m^2 \mid,
\end{equation}
where $V(\phi,H)=V^{(0)}+V^{(1)}+ \cdots$ is the expansion in the number of 
loops and $\delta(0)$ is the vacuum value of the scalar field determined in  
a tree approximation.

It is convenient for what follows to introduce the dimensionless quantities:
$h=H/H_0 (H_0=M^2_w/e),\phi=\phi_c/\delta(0), K =m_H^2/M_w^2,$ $B=\beta 
M_w, 
\tau=1/B = T/M_w,$$ {\cal V}= V/H^2_0$ and $M_w = \frac{g}{2}\delta(0)$.

Explicit forms of $V^{(1)}$ at zero temperature are quoted in Refs.\cite{Sk2},
\cite{DTZ}.
Ommiting details of  computations which can be found in Refs. \cite{SVZ}, 
\cite{Cab}, \cite{Sk2}, we present the finite temperature contributions 
of boson fields in the form \cite{Rez}:
\begin{equation} \label{14a}  Re \omega^{(1)}_w = - 4 \frac{\alpha}{\pi}
\frac{h}{B} ( 3\omega_0 + \omega_1 - \omega_2 ),
\end{equation}
where
\begin{equation} \omega_0 = \sum\limits_{p=0}^{\infty} \sum\limits_{n=1}^
{\infty} \frac{x_p}{n} K_1(nBx_p) ,~ x_p = (\phi^2 + h +2ph )^{1/2};
\end{equation}
\nonumber\\
\begin{equation} \omega_1 = \sum\limits_{n=1}^{\infty} \frac{y}{n} K_1(nBy),
~ y = (\phi^2 - h )^{1/2}
\end{equation}
\nonumber\\
and in the range of parameters $ \phi^2 < h $ after analytic continuation
\begin{equation} Re \omega_1 = -\frac{\pi}{2} \sum\limits_{n=1}^{\infty}
\frac{\mid y \mid}{n} Y_1(nB\mid y \mid) ,
\end{equation}
\nonumber\\
\begin{equation} \omega_2 = \sum\limits_{n=1}^{\infty} \frac{z}{n} K_1(nBz),
z = (\phi^2 + h )^{1/2},
\end {equation}
and $K_n(x), Y_n(x)$ are the Bessel functions. 
The imaginary part of $\omega_{1}$ will be cancelled by  the contribution of  ring 
diagrams with the tachyonic mode. So,  it is  not adduced here.

The fermion finite temperature contribution can be written in the form:
\begin{eqnarray} \label{20} \omega_{f}&=&  4 \frac{\alpha}{\pi}\sum
\limits_{f}\Bigl\{\sum_{p=0}^{\infty}\sum_{n=1}^{\infty}(-1)^n \Bigl[\frac{
(2ph + K_f\phi^2)^{1/2} h}{Bn} K_1((2ph + K_f\phi^2)^{1/2}Bn)
\nonumber\\
&+& \frac{(2p+2)h + K_f\phi^2)^{1/2}}{Bn} h
K_1(((2p+2)h + K_f\phi^2)^{1/2}Bn)\Bigr]\Bigr\}
\end{eqnarray}
where $K_f = m_f^2 /M_w^2 $. The above expressions will be used in the numerical investigations 
of symmetry behaviour.
~\\~

   3.    It was shown by Carrington \cite{Car} that at $T\not = 0$ a 
consistent calculation of the EP based on generalized propagators, which 
include the polarization operator insertions, requires the ring
diagrams to be added simultaneously with the one-loop terms.
These diagrams  essentially affect  the phase transition at high temperatre 
and zero field \cite{Tak},\cite{Car}. Their importance at $ T$ and $H \not= 
0$ was also pointed out in literature \cite{SVZ} but, as far as we know, 
this part of the EP has not been calculated, yet.

As is known \cite{Tak}, the sum of ring diagrams describes a dominant 
contribution of long distances. It gives significant effect when 
massless states appear in a system. So, this type of diagrams has to be
calculated when a symmetry restoration is investigated. Now, let us turn to 
computations of $V_{ring}(H,T)$. It is described by the standard 
expression \cite{Tak},\cite{Car},\cite{SVZ}:
\begin{equation} \label{21}
V_{ring} = - \frac{1}{12\pi\beta} \{Tr[M^2(\phi) +
\Pi_{00}(0)]^{3/2} - M^3(\phi)\},
\end{equation}
where the trace means summation over  the all  contributing states, $M(\phi)$ 
is the tree mass of the  corresponding state and $\Pi_{00}(0) = \Pi(k=0,T,H)$ for 
the Higgs particle and $\Pi_{00}(0)=\Pi_{00}(k=0,T,H)$ are the zero-zero 
components of the polarization operators in a magnetic field taken at 
zero momenta.
The above contribution has order $ \sim g^3 (\lambda^{3/2}) $ in coupling 
constant whereas the two-loop terms are of order $\sim g^4,\lambda^2 $. As 
$\Pi_{00}(0)$ the high temperature limits of polarization functions have 
to be substituted which are of orders $\sim T^2$ for leading terms and 
$\sim g\phi_c T,~ (gH)^{1/2}T~ ( \phi_c/T << 1, (gH)^{1/2}/T << 1)$ for subleading ones.

For the next step of calculations, we remind that the EP 
is the generating functional of the one-particle irreducible Green functions 
at zero external momenta. So, to have $\Pi(0)$ we may  just calculate the
second  derivative with respect to $\phi$ of the potential $V^{(1)}(H,T,\phi)
$ in the  limit of high temperature, $T >>\phi, T >> (eH)^{1/2}$ and set $\phi = 0$. 
This limit can be calculated by means of the Mellin  transformation technique 
and the result looks as  follows:
\begin{eqnarray} \label{22} V^{(1)}(H,&\phi&,T \rightarrow \infty) = \left.[\Bigl(
\frac{C_f}{6}\phi^2_c + \frac{\alpha\pi}{2 cos^2\theta_w}\phi^2_c + \frac{g^2}
{16}\phi^2_c \Bigr) T^2 \right.]
\nonumber\\
&+& \left.[ \frac{\alpha \pi}{6} (3\lambda\phi^2_c - \delta^2(0))T^2 - \frac{
\alpha}{cos^3\theta} \phi^3 T - \frac{\alpha}{3} (\frac{3\lambda\phi^2_c -
\delta^2(0)}{2})^{3/2} T \right.]\\
&-& \frac{1}{2\pi} (\frac{1}{4}\phi^2_c + gH)^{3/2} T + \frac{1}{4\pi} eH T
(\frac{1}{4}\phi^2_c + eH )^{1/2} + \frac{1}{2} eH T (\frac{1}{4} \phi^2_c -
eH )^{1/2}, \nonumber
\end{eqnarray}
The parameter $C_f = \sum\limits_{i=1}^{3} G^2_{il} + 3\sum\limits_{i=1}^{3} 
G^2_
{iq}$ determines the fermion contribution of the order $\sim T^2$ having  
relevance to our problem. We also omitted $\sim T^4$ terms in the EP. 
The terms  of the type $\sim log [T/f(\phi,H)]$ cancel the logarithmic terms in 
the temperature independent parts. Considering the high temperature 
limit we restrict ourselves by the linear and quadratic in $T$  terms, only.

Now, differentiating this expression twice with respect to $\phi$ and setting 
then $\phi=0$, we obtain
\begin{eqnarray} \label{23} \Pi_{\phi}(0) &=& \left.[\frac{\partial^2 V^{(1)}
(\phi,H,T)}{\partial \phi^2} \mid_{\phi=0} \right.]
\nonumber\\
&=& \frac{1}{24\beta^2}\Bigl( 6\lambda + \frac{6 e^2}{\sin^2 2\theta_w}
+ \frac{3 e^2}{\sin^2 \theta_w} \Bigr)\nonumber \\
  &+& \frac{2\alpha}{\pi} \sum\limits_{f}\left.[ \frac{\pi^2 K_f}{3\beta^2} - \mid q_f H \mid K_f \right.]
\nonumber\\
&+& \frac{(eH)^{1/2}}{8\pi \sin^2\theta_{w}\beta} e^2 (3\sqrt{2} \zeta(-\frac{1}
{2},\frac{1}{2}) - 1 ).
\end{eqnarray}
Here, $K_f = m^2_f/M^2_w$ and $q_f$ is the electric charge of the fermion. We also have added the fermion H-dependent contribution (second term of the fermion sum) which eppears in the one-loop EP.
The terms $\sim T^2$ in Eq.(\ref{23}) give the known temperature mass squared coming 
from the boson and fermion sectors. The last $H$-dependent term 
is negative since the difference in the brackets is $3\sqrt{2}\zeta (-\frac{1}
{2},\frac{1}{2}) - 1 \simeq - 0,39$. Formally, this term has to produce an 
instability for strong fields but actually it happens for $(eH)^{1/2} >> T$ 
when the asymptotic series is not applicable. Substituting expression 
(\ref{23}) into Eq.(\ref{21}) we  obtain (in the dimensionless  variables),
\begin{equation} \label{24} {\cal V}^{\phi}_{ring} = - \frac{1}{12B}
\Bigl\{(\frac{3\phi^2 - 1}{2} K  + \Pi_{\phi}(0) \Bigr\}^{3/2} + \frac{\alpha}
{3B} K^{3/2} (\frac{3 \phi^2 - 1}{2})^{3/2}.
\end{equation}

To find the $H$-dependent Debye masses of photons and $Z$-bosons the following 
procedure will be used. First, we calculate the one-loop EP of the $W$-bosons 
and fermions in a magnetic field and some "chemical potential", $\mu$, which 
plays the role of the auxiliary parameter. Then, by differentiating them twice 
with respect to $\mu$ and setting $\mu = 0$ the mass squared $m^2_D$ will be 
found. Let us describe that in more detail for the case of fermions  where the result is known.

The temperature dependent part of the one-loop EP of constant magnetic field and 
non-zero chemical potential induced by an electron-positron vacuum polarization 
is \cite{Elm}:
\begin{equation} \label{25} V^{(1)}_{ferm.} =
  \frac{1}{4\pi^2} \sum\limits_{l-1}^{\infty}
(-1)^{l+1}\int\limits_{0}^{\infty} \frac{ds}{s^3} exp(\frac{-\beta^2 l^2}
{4s} - m^2s ) eHs coth(eHs) cosh(\beta l\mu),
\end{equation}
where $m$ is the electron mass, $e = g sin \theta_w$ is electric charge and 
proper-time representation is used. Its second derivative with respect to $\mu$  
taken at $\mu = 0$ can be written in the form
\begin{equation} \label{26}  \frac{\partial^2 V^{(1)}_{ferm.}}{\partial \mu^2}=
\frac{eH}{\pi^2}\beta^2 \frac{\partial}{\beta^2}\sum\limits_{l=1}^{\infty}
(-1)^{l+1}\int\limits_{0}^{\infty} \frac{ds}{s} exp(-m^2s -\beta^2 l^2/4s)
coth(eHs).
\end{equation}
Expanding $coth (eHs) $ in series and integrating over $s$ we obtain in
the limit of $ T >> m,~T >> (eH)^{1/2}$:
\begin{equation} \label{27} \sum\limits_{l=1}^{\infty} (-1)^{l+1} [\frac{8m}
{\beta l} K_1(m\beta l) + \frac{2}{3}\frac{(eH)^2 l\beta}{m} K_1(m\beta l)+
\cdots ]
\end{equation}
Series in $l$ can easily be calculated by means of the Mellin transformation 
(see, for example, Refs.\cite{SVZ}). To have the Debye mass squared it is 
necessary to differentiate Eq.(\ref{27}) with respect to $\beta^2$ and to take 
into  account the relation $\mu \rightarrow ieA_0$ \cite{SVZ}  of  the parameter $\mu$ with the zero component of 
the electromagnetic potential. In this way we obtain  finally
\begin{equation} \label{28} m^2_{D} = g^2 sin^2\theta_w ( \frac{T^2}{3} -
\frac{1}{2\pi^2} m^2 + O((m\beta)^2, (eH\beta^2)^2 ) ).
\end{equation}
This is the well known result calculated from the photon polarization operator (see 
for example \cite{VZM}).
As one can see, the dependence on $H$ appears in the order  $\sim T^{-2}$.

Now, let us apply this procedure for the case of the $W$-boson contribution. The  
corresponding EP
(temperature dependent part) calculated at non-zero $T,\mu$ 
looks as follows,
\begin{equation} \label{29} V^{(1)}_w = - \frac{eH}{8\pi^2}\sum\limits_{l=1}^{
\infty}\int\limits_{0}^{\infty} \frac{ds}{s^2}exp(-m^2 s -l^2\beta^2/4s)[\frac{
3}{sinh(eHs)}+ 4 sinh(eHs)] cosh(\beta l\mu).
\end{equation}
All the notations are obvious. The first term in the squared brackets gives the 
triple contribution of the charged scalar field and the second one is due to 
the interaction with the $W$-boson magnetic moment. Again, after  differentiation 
twice with respect to $\mu$ and setting $\mu = 0$ it can be written as
\begin{equation} \label{30} \frac{\partial^2 V^{(1)}_w}{\partial \mu^2} =
\frac{eH}{2\pi^2}\beta^2\frac{\partial}{\partial\beta^2} \sum\limits_{l=1}^
{\infty}\int\limits_{0}^{\infty}\frac{ds}{
s}exp(- m^2s - \frac{l^2\beta^2}{4s})[\frac{3}{sinh(eHs)} + 4
sinh(eHs)].
\end{equation}
Expanding $sinh^{-1}s$ in series over Bernoulli's polynomials and carrying out 
all the calculations described above, we obtain for the $W$-boson contribution 
to $m^2_D$ of the electromagnetic field,
\begin{equation}
\label{31} m^2_D = 3 g^2 sin^2 \theta_w [ \frac{1}{3} T^2 -
\frac{1}{2\pi} T(m^2 + g sin\theta_w H)^{1/2} -
\frac{1}{8\pi^2}(g sin\theta_w H)].
\end{equation}
As before, it is necessary to express masses through the vacuum value of the 
scalar condensate $\phi_{c}$. In the same way the $Z$-boson part $V^{z}_{ ring}$ 
can be calculated. The only difference is the additional mass term of $Z$-field 
and an additional term in the Debye mass due to  the neutral
current $\sim \bar{\nu}\gamma_{\mu}\nu Z_{mu}$ . These three fields - $\phi, 
\gamma, Z$ , - which becomes massless in the restored phase, contribute into 
$V_{ring}(H,T)$ in an external magnetic field. At zero field there is 
also a term due to the $W$-boson loops. But when $H \not = 0$ the charged 
particles acquire $\sim eH$  masses. The corresponding fields remain short-range 
ones in the restored phase of the vacuum and therefore do not contribute. 

A separate consideration should be spared to the tachyonic (unstable)  mode in the 
$W$-boson spectrum: $p^2_0 = p^2_3 + M^2_w - eH$. In the fields $eH \sim 
M^2_w$ the  mode becomes a long range state. Therefore,   it has to be included in 
$V_{ring}(H,T)$ side by side  with  other considered neutral fields.  But in this case
it is impossible  to take advantage of formula (\ref{21}). So, we turn to the  EP written in terms of the generalized propagators. 

For our purpose  it will be convenient to use the expression for the  generalized EP written 
as the sum over the modes in external magnetic field \cite{SVZ}:
\begin{equation} \label{TDVZ}  V^{(1)}_{gen} =  \frac{eH}{2\pi \beta} \sum \limits_{l= 
-\infty}^{+ \infty} \int\limits_{- \infty}^{+ \infty} \frac{dp_3}{2\pi} \sum\limits_{n = 0, 
\sigma = 0,\pm 1}^{\infty} log [\beta^2(\omega^2_l + \epsilon ^2_{n,\sigma,p_3} + 
\Pi(T,H) )] ,
\end{equation}
where $\omega_l = \frac{2\pi l}{\beta}$ ,  $\epsilon^2_n = p_3^2 + M^2_w + (2n + 1 - 
2\sigma) eH $ and $\Pi(H,T)$ is the Debye mass of  $W$-bosons in a magnetic field.
Denoting   as $D^{- 1}_0(p_3,H.T)$ the sum $ \omega^2_l + \epsilon^2$, one can rewrite eq.
(\ref{TDVZ}) as follows:
\begin{eqnarray}
\label{TDVZ1}
V^{(1)}_{gen} &=& \frac{eH}{2\pi\beta}
\sum\limits_{-\infty}^{+\infty}\int\limits_{-\infty}^{+\infty}
\frac{d p_3}{2\pi}
\sum\limits_{n,\sigma} log[\beta^2 D^{- 1}_0(p_3,H,T)]
\nonumber\\
&+&  \frac{eH}{2\pi\beta} \sum\limits_{- \infty}^{+ \infty}
\int\limits_{- \infty}^{+\infty}
\frac{d p_3}{2\pi} \{ log[ 1 + ( \omega^2_l + p^2_3 + M^2_w - eH)^{-1}
\Pi(H,T) ] \nonumber\\
&+&
\sum\limits_{n \not = 0, \sigma \not = +1 }
log[ 1 +  D_0 ( \epsilon^2_n , H,T) \Pi(H,T) ] \}.
\end{eqnarray}
Here,  the first term  is just the one-loop contribution of $W$-bosons,  the second one 
gives the  sum of ring diagrams  of the unstable mode ( as it can easily be verified by 
expanding the logarithm into a series). The last term describes the sum of the short range 
modes and has to be omitted. 

Thus, to find  $V^{unstable}_{ring}$ one must calculate the second term in Eq. (\ref{TDVZ1}). In the high temperature limit we obtain:
\begin{equation} \label {Vunst}
V^{unstable}_{ring} = \frac{eH}{2\pi\beta} \{ ( M^2_w
- eH + \Pi (H, T) )^{1/2} - ( M^2_w - eH )^{1/2} \}.
\end{equation}
By summing up  the one-loop EP and  all the  terms $ V_{ring}$ , we arrive at the total 
consistent in leading order EP. 

Let us note the most important featurers of the above expression. It is seen that the last term 
in Eq.(\ref{Vunst}) exactly cancels the ``dangerous'' becoming imaginary term in the 
$V^{(1)}$ Eq. (\ref{22}). So, no instabilities appear at sufficently high temperatures when $\Pi ( H,T) > 
M^2_w - eH $ and the EP is real. To make a quantitative estimate of the range of validity of 
the total EP
it is necessary to calculate the mass operator of $W$-boson in a magnetic field at finite 
temperature and hence to find $\Pi(H,T)$. This is separate  and enough complicated
problem which is not considered in detail, here. The high temperature limit of the polarization function has been calculated in Ref.\cite{SkSt} and the result looks as follows: $Re \Pi(H,T) = 26,96 \frac{e^2}{4\pi}(eH)^{1/2}T$.
\footnote{This result disagrees with the corresponding one of Refs.\cite{Elmp},\cite{Pers}where this value was found to be zero. Most probably, the discrepancy is because of the calculation procedure used by these authors. They have calculated the polarization operator of the charged gauge field at $H = 0$ and then have averaged the result in the state $n = 0, \sigma = + 1$ calling this ``weak field approximation''. Our expresion is the high temperature limit of the mass operator which takes into account the external field exactly. }  
  Below, the obtained mass will be  
 substituted into  $V^{unstable}_{ring}$ and 
used in the following estimations.

~\\~

 4. Now, we are going to investigate the symmetry behaviour at high temperature 
and strong magnetic fields. In order to do that we shall consider the function $ 
{\cal V^{'}} =Re[ {\cal V}(h,\beta,\phi) - {\cal V}(h,\beta,0)] $ describing the 
symmetry restoration. In fact, we have observed two standard types of that. For weak fields the minimum position of the EP $\phi_{min}(h, B)$ is decreased smoothly
from unit at low temperature to zero with temperature increasing. That is typical for the second order phase transition. For strong fields $h > 0.1$ that corresponds to $H > 0.1\cdot 10^{24} G$ the phase transition becomes of the first order. In this case it is important to check whether the instability threshold $ H_0 = M_w^2 / e$ is desposed in the local minimum of the EP for the  field strengths, when the phase transition of the first order happens. That is, whether the effective mass squared $M^2_w(H, T, \phi_c(H,T)) = \frac{g^2}{4}\phi^2_c(H,T_c) - eH + \Pi(H,T_c)$ is positive or not for the field strengths applied. If this is the case, the calculation is self-consistent and the magnetic field is stable otherwise the evolution of the unstable mode should be considered. The investigation to be of the type as has been carried out in Refs. \cite{Sk1}, \cite{AO}, \cite{MDT}.   

Next what is necessary to note is the condition which determines the temperature of the phase transition $T_c$.  The transition happens for the case
\begin{equation} \label{phtrH}
V_{restored}(H, T_c, 0) = V_{broken}(H, T_c, \phi_c(H, T_c))
\end{equation}
when the depth of the minima located at the begining, $\phi_c = 0$, and at $ \phi_c=\phi_{min} \neq 0$ is coinsiding. 

 In Table 1 we present the results of numeric investigation of the first order phase transition for strong fields $h$. We collected the characteristics of it which are the  most important for the baryogenesis problem.

  ~\\~

\begin{tabular}{c c c c c c}\hline
~\\~
~h&    K &  $T_c(GeV)$  & $\phi(h,T_c)$& $R$  &  $a$ \\[5pt] 
\hline
~1.5&  0.85 & 105.66    & 0.44385   & 1.03338 & 2.77958  \\
~1.5&  1.25 & 122.36    & 0.38987   & 0.78381 & 2.07133  \\
~1.5&   2   & 146.99    & 0.34496   & 0.57732 & 1.43536 \\ \hline
~2& 0.85 & 105.12  & 0.45717 & 1.06986  & 3.74226       \\
~2& 1.25 & 120.56  & 0.44497 & 0.90795  & 2.84485       \\
~2&  2   & 146.04  & 0.37815 & 0.63698  & 1.92293      \\ \hline
~\\~
Table 1.  
\end{tabular}

The second column shows the values of the parameter $K = m^2_H/M^2_w$, corresponding to the Higgs boson masses $75 GeV, 90 GeV$ and $115 GeV$, respectively. The third one gives the critical temperatures. The fourth column determines the local minimum position (in dimensionless units). It gives possibility to find the mass squared of the W-bosons, $M^2_w(H,T) = (81 GeV)^2 \phi^2,$ and the jump of the order parameter $\phi_c(H,T_c) = (246 GeV)\phi(H, T_c)$. Hence the ratio $R = \phi_c(H, T)/T_c$ describing the the advantage of the baryogenesis \cite{RS} can be calculated. In the last column we show the ratio $a = H/T_c^2$. It can be used to compare our results with that of other papers.
   
As one can see, the increase in $h$ makes the phase transition of the first order stronger. However, even for the masses $m_H \sim 75 GeV$ the condition $R > 1.2 -1.5$ necessary for the baryogenesis \cite{Shap2}, \cite{RS} is not satisfied. We also note that for strong fields (because of small $\phi(H,T)$) the effective W-boson mass squared is negative. That means instability of the local minimum. However, we belive that the presence of the W- and Z-boson condensates does not increase the R value. 
 Thus, our analysis shown that for the case of external magnetic field the baryogenesis does not survive in the minimul SM. 

Let us compare the results presented with the ones in Refs.\cite{Shap1} - \cite{Lain}. Remind that these authors have considered the external hypermagnetic field and restricted themselves by its the tree level effects. Besides, they have taken into account the contribution of the t-quark, only. Because of these circumstances, they, actually, omitted the most important point that at high temperatures and strong fields the light fermions are important and the local minimum position $\phi_{min}(H,T)$ is small at the restoring tempereture $T_c$. Just the latter point significantly decreases the value of the parameter R in the magnetic field.  Since in the broken phase the magnetic and the hypermagnetic fields coincide, our investigation is strightforwardly relevant to the latter case in the respect of the EP curve. Second, in Ref.\cite{Elm1} it was determined that the ratio $a(H,T) = H/T^2$ should be of order $\sim 0.1 - 0.5$. This estimate has then been used in the lattice calculations in Refs.\cite{Shap2}, \cite{Lain}. In our calculations for strong magnetic fields the values $a(H,T)$ of larger order have been determined.
We would like to complete with other remark concerning the comparison with the hypermagnetic field. Actually, the main mathematical difference between these cases consists in the conditions determining the temperature $T_c.$ In the hypermagnetic field the condition (\ref{phtrH}) must be replaced by the one taking into account the partial screening of the external field in the broken phase (see Refs. \cite{Shap1}, \cite{Elm1}). Besides, in the restored phase the W-bosond do not interact with the external field, therefore no instabilities occure. We will conside this case  in more detail in other publication.   

The authors thank Alexei Batrachenko and Vadim Demchik for they help in  carring out of numeric calculations.

\end{document}